\documentclass[aps,prl,twocolumn]{revtex4}%
\usepackage{amsfonts}
\usepackage{amsmath}
\usepackage{amssymb}

  \usepackage[pdftex]{graphics}
  \DeclareGraphicsExtensions{.pdf}

\begin{document}
\title{Efficient evaluation of accuracy of molecular quantum dynamics using dephasing representation}
\author{Baiqing Li}
\author{Cesare Mollica}
\author{Ji\v{r}\'{\i} Van\'{\i}\v{c}ek}
\email{jiri.vanicek@epfl.ch}
\affiliation{Laboratory of Theoretical Physical Chemistry, Institut des Sciences et
Ing\'{e}nierie Chimiques, \'{E}cole Polytechnique F\'{e}d\'{e}rale de Lausanne
(EPFL), CH-1015, Lausanne, Switzerland}
\keywords{quantum dynamics, semiclassical, quantum fidelity, Loschmidt echo, dephasing,
ab initio method}
\pacs{PACS number}

\begin{abstract}
Ab initio methods for electronic structure of molecules have reached a
satisfactory accuracy for calculation of static properties, but remain too
expensive for quantum dynamical calculations. We propose an efficient
semiclassical method for evaluating the accuracy of a lower level quantum dynamics, as
compared to a higher level quantum dynamics, without having to perform any 
quantum dynamics. The method
is based on dephasing representation of quantum fidelity and its feasibility
is demonstrated on the photodissociation dynamics of CO$_{2}$. We suggest how
to implement the method in existing molecular dynamics codes and describe a
simple test of its applicability.
\end{abstract}

\date{\today}

\maketitle

Ab initio methods for electronic structure of molecules have reached a
satisfactory accuracy for calculation of static properties, such as energy
barriers or force constants at local minima of the potential energy surface
(PES). The most accurate of such methods remain out of reach when one wants to
describe molecular properties depending on the full quantum dynamics
\cite{clary:2008}. To make a calculation feasible, one has to approximate
the dynamics of the system \cite{makri:1999,miller:2005} or the PES
\cite{clary:2008}, but both approaches can have nontrivial effects on the
result \cite{rabitz:1987}. We consider only the second approach, in
which quantum dynamics is done exactly but on a PES obtained by a lower level
electronic structure method that is less accurate but also less expensive.
When such a calculation is finished, its accuracy is not known because of the
forbidding expense of the dynamics on the more accurate potential. In this
Letter, we propose an efficient and accurate semiclassical (SC) method
for evaluating the accuracy of the lower level quantum dynamics without having
to perform the higher level quantum dynamics. Since our method does not even
require computing the lower level quantum
dynamics, it can also be used to justify, in advance, investing computational
resources into the lower level calculations.

For simplicity, we use the Born-Oppenheimer
approximation and focus on the quantum dynamics of nuclei, although our
method applies to any quantum dynamics and should be valid even when nonadiabatic
effects are important. The time-dependent Schr\"{o}dinger equation is solved
in two stages: First, the time-independent equation for
electrons is solved with fixed nuclear configurations, and then nuclear motion
is calculated on the resulting electronic PES. We
consider three PESs: $V_{\text{exact}}$ is the
\emph{exact} PES that describes our system. $V_{\text{acc}}$ is a very
\emph{accurate} high-level electronic structure PES (presumably
\textquotedblleft almost exact\textquotedblright), which is too expensive to
be used for quantum dynamics. $V_{\text{appr}}$ is an \emph{approximate} PES,
obtained by a lower-level electronic structure method (or by an analytical fit
of $V_{\text{acc}}$) and ``cheap'' enough
 to be used for quantum dynamics.

Various quantities can describe different time-dependent features of a quantum
system, but a single quantity that includes \emph{all} information about the
system is the time-dependent wave function, $\psi(t)$. One way to evaluate the
accuracy of quantum dynamics on the approximate PES would therefore be to
compute the quantum-mechanical (QM) overlap $f_{\text{QM}}(t):=\langle
\psi_{\text{exact}}(t)|\psi_{\text{appr}}(t)\rangle$, where the subscript of
$\psi$ denotes the corresponding PES used for propagating the initial
state $\psi(0)$. The quantity $F(t):=\left\vert f_{\text{QM}}(t)\right\vert
^{2}$ is known as \emph{quantum fidelity} or \emph{Loschmidt echo}, and has
been defined by Peres \cite{peres:1984} to measure the sensitivity of quantum
dynamics to perturbations. Much effort has been devoted to the study
of temporal decay of fidelity and many universal regimes have been found \cite{gorin:2006}. In our setting,
if $F(t)\approx1$ for all times up to $t_{\text{max}}$, we can trust quantum
dynamics on the approximate potential $V_{\text{appr}}$ and use the resulting
$\psi_{\text{appr}}(t)$ to compute all dynamical properties up to
$t_{\text{max}}$. In calculations, we do not know $V_{\text{exact}}$
and must use the accurate potential $V_{\text{acc}}$,
and so we approximate $f_{\text{QM}}$ by
\begin{equation}
f_{\text{QM}}:=\left\langle \psi_{\text{acc}}(t)|\psi_{\text{appr}%
}(t)\right\rangle .\label{fidQM}%
\end{equation}
Since $V_{\text{acc}}$ is too expensive, $\psi_{\text{acc}}(t)$ cannot be computed. We describe a method which
gives an accurate estimate of $f_{\text{QM}}$ without having to compute
$\psi_{\text{acc}}(t)$ nor $\psi_{\text{appr}}(t)$.

The method is based on the \emph{dephasing representation} (DR) of quantum
fidelity, a SC approximation proposed by one of us to evaluate
fidelity in chaotic, integrable, and mixed systems even in nonuniversal
regimes sensitive to the initial state and details of dynamics
\cite{vanicek:2004a,vanicek:2006}. Presently, we are interested in a specific
type of \textquotedblleft
perturbation,\textquotedblright\ namely the difference $\Delta
V=V_{\text{appr}}-V_{\text{acc}}$ between the approximate and accurate PESs.
The DR of fidelity amplitude is an interference integral%
\begin{align}
f_{\text{DR}}(t)  &  :=\int dx^{0}\,\rho_{\text{W}}(x^{0})\exp\left[  -i\Delta
S(x^{0},t)/\hbar\right]  ,\label{fidDR}\\
\Delta S(x^{0},t)  &  =\int_{0}^{t}d\tau\,\Delta V\left[  q_{\text{acc}}%
^{\tau}(x^{0})\right]  . \label{actP}%
\end{align}
Here $x$ denotes a point $\left(  q,p\right)  $ in phase space, the superscript
is the corresponding time, $\Delta S(x^{0},t)$ is the action due to
$\Delta V$ along the trajectory $q_{\text{acc}}^{t}(x^{0})$ of $V_{\text{acc}%
}$, and $\rho_{\text{W}}$ is the Wigner function of the initial state $\psi$,%
\[
\rho_{\text{W}}(x)=h^{-d}\int d\xi\mathbf{\,}\psi^{\ast}\left(  q+\frac{\xi
}{2}\right)  \psi\left(  q-\frac{\xi}{2}\right)  \,\exp\left(  i\frac{\xi\cdot
p}{\hbar}\right)  .
\]
In \textquotedblleft dephasing representation,\textquotedblright\ all of
fidelity decay appears to be due to interference and none due to decay of
classical overlaps.

Surprising accuracy of the DR was justified by the shadowing theorem
\cite{vanicek:2004a}, or, in the case of the initial state $\psi (0)$ supported by
a Lagrangian manifold, by the structural stability of manifolds
\cite{cerruti:2002}. Interestingly, validity of the DR goes much beyond
validity of the SC approximation for quantum dynamics on
$V_{\text{acc}}$ or $V_{\text{appr}}$. Qualitatively, this is due to mitigating the \textquotedblleft sign problem\textquotedblright\ of
quantum or SC dynamics: large actions needed for dynamics on
$V_{\text{acc}}$ or $V_{\text{appr}}$ are replaced by much smaller actions
$\Delta S$ needed for fidelity calculation. Rapid oscillations in the
SC expression for the dynamics on $V_{\text{acc}}$ or $V_{\text{appr}}$
are replaced by much slower oscillations in the DR. In a chaotic system where
$\sim10^{35}$ trajectories would be needed for computing $\psi_{\text{acc}%
}(t)$ semiclassically, as few as $1000$ trajectories were sufficient to
compute fidelity amplitude \cite{vanicek:2003a}. Accuracy of DR was explored numerically in Refs.
\cite{vanicek:2004a,vanicek:2004b,vanicek:2006} which suggest that
$F_{\text{DR}}$ starts to deviate from $F_{\text{QM}}$ after the Heisenberg
time $t_{H}=\hbar/\Delta E$ where $\Delta E$ is the mean level spacing. Errors of DR can be estimated analytically, suggesting that DR breaks
down for very large perturbations, when the effective perturbation is
larger than the square root of the effective Planck's constant
\cite{vanicek:2009}. However, DR remains accurate for fairly large
perturbations, even when corresponding classical trajectories of
$V_{\text{acc}}$ and $V_{\text{appr}}$ are completely different
\cite{vanicek:2004a,vanicek:2004b,vanicek:2006}.

The fundamental reason why quantum dynamics calculations are expensive is
nonlocality of quantum mechanics: Wave function $\psi(t+\Delta t)$ at any
point in space depends in general on $\psi(t)$ in the whole space. There are
many computational methods for quantum dynamics, but for the sake of
demonstration, we consider two methods that represent two very different
general approaches.

The first approach starts with the construction of a global PES, with a
computational cost $cn^{d}$ where $c$ is the cost of a single potential
evaluation, $d$ is the number of degrees of freedom (DOF), and $n$ is the
number of grid points in each DOF. Once the PES is known, dynamics can be
performed, e.g., by the split-operator method \cite{feit:1982}. In this
method, the quantum
evolution operator for time step $\Delta t$ is approximated by%
\[
e^{-iH\Delta t/\hbar}=e^{-iT\Delta t/\hbar}e^{-iV\Delta t/\hbar}+O(\Delta
t^{2}),
\]
where $H=T+V$ is the Hamiltonian of the system and $T$ is the kinetic energy
operator. Quantum dynamics consists of alternate kinetic and potential
propagations (which are just multiplications in momentum and coordinate
representations, respectively)
and a fast Fourier transform (FFT) to switch the representation in between.
The complexity of FFT is $O(N\log N)$ where $N=n^{d}$ denotes the dimension of the
Hilbert space, so the cost of propagation is $O(d\,t\,n^{d}\log n)$. 
Note that the same number of potential energy
evaluations is required no matter how long the propagation is. This becomes an advantage for long-time dynamics
calculations and a disadvantage for short-time ones.
Finally, memory requirements make this approach
feasible only for very small systems.

In the second approach, potential energy is calculated \textquotedblleft on
the fly\textquotedblright\ only in the vicinity of the propagated wave packet.
At each time step, the cost is $c\,\tilde{n}^{d}$ where $\tilde{n}$ is the
number of grid points in each DOF on which $\psi(t)$ is not negligible.
Presumably, $\tilde{n}\ll n$, but the exponential scaling with $d$ remains.
Assuming that $\log\tilde{n}<c$, the cost of actual propagation (e.g., by FFT)
is negligible to the cost of potential evaluation, and the overall cost of the
dynamics is $O(c\,\tilde{n}^{d}\,t)$.

In the DR, potential energy and forces can also be evaluated on the fly, along
classical trajectories. But at each time step, the cost is only
$O(d\,c\,n_{\text{paths}})$ where $n_{\text{paths}}$ is the number of
classical trajectories used. 
The ``hidden cost'' in usual SC
approximations is the strong dependence of $n_{\text{paths}}$ on $t$, $d$, or
the type of dynamics. In Ref. [8], it was shown rigorously that
$n_{\text{paths}}=C(F_{\text{DR}})\sigma^{-2}$ where $F_{\text{DR}}$ is the
value of fidelity one wants to simulate, $\sigma_{F_{\text{DR}}}$ is the error
(due to finite $n_{\text{paths}}$) that one wants to reach, and $0 \leq C \leq
3$. 
Consequently, for given $F_{\text{DR}}$ and $\sigma$, the required number
of trajectories is independent of $t$, $d$, or the type of dynamics!
It was also shown that when $F \rightarrow 1$,  which is most interesting in
our application, $C \rightarrow 0$, i.e., $n_{\text{paths}}$ needed for
convergence of $F_{\text{DR}}$ becomes even smaller. The overall cost of the
DR dynamics is $O(d\,c\,n_{\text{paths}}\,t)$. In particular, there is no
exponential scaling with the number of DOF or time.

Clearly, DR is faster than the construction of a global PES and than the
quantum dynamics on the fly. Only at very long times $t$, in the first quantum
approach, since the global PES is already constructed, the cost of propagation
becomes dominant and one would expect that QM dynamics would beat DR dynamics
which requires new potential evaluations. But even then the ratio of the costs
of DR and QM is $c\,n_{\text{paths}}/n^{d}\log n$. Assuming that the number
of nuclear DOF is comparable to the number of electronic DOF, then even for
the most accurate ab initio methods (e.g., the coupled clusters) $c$
scales polynomially with $d$, and so for
large enough $d$, DR will still be faster than QM dynamics. 

The DR calculation can be further accelerated by exchanging roles of
$V_{\text{acc}}$ and $V_{\text{appr}}$. We will denote the DR expression
defined in Eqs. (\ref{fidDR})-(\ref{actP}) by DR1 and by DR2 an analogous
expression,
\begin{align}
f_{\text{DR2}}(t) &  :=\int dx^{0}\,\rho_{\text{W}}(x^{0})\exp\left[  i\Delta
S(x^{0},t)/\hbar\right]  ,\label{fidDR2}\\
\Delta S(x^{0},t) &  =\int_{0}^{t}d\tau\,\Delta V\left[  q_{\text{appr}}%
^{\tau}(x^{0})\right]. \label{actP2}%
\end{align}
DR2 denotes the DR
computed as an interference integral due to action of $-\Delta V$ along the
trajectory of $V_{\text{appr}}$. From definition (\ref{fidQM}), it is clear that exchanging $V_{\text{acc}}$ and
$V_{\text{appr}}$ results in complex conjugation of $f_{\text{QM}}$ and has no
effect on $F_{\text{QM}}$. In principle, there could be an effect on
$f_{\text{DR}}$ because expression (\ref{fidDR}) does not have such symmetry,
but numerical evidence presented below shows that DR1 $\approx$\ DR2,
providing further support for the approximation. 
In applications where efficiency is important, one should choose DR2 over DR1 since
DR2 requires values and gradients of the \textquotedblleft
cheaper\textquotedblright\ PES ($V_{\text{appr}}$) but only values of the more
expensive PES ($V_{\text{acc}}$) whereas DR1 requires values of
$V_{\text{appr}}$ but both values and gradients of $V_{\text{acc}}$. In
applications where calculation of DR1 is affordable, comparison of DR1 and DR2
results can be used as a validity test of the DR method since comparison
with $f_{\text{QM}}$ will not be available (computation of $f_{\text{QM}}$
would require full quantum dynamics on $V_{\text{acc}}$). A large difference between DR1 and DR2 results would be a sign of the
breakdown of DR. So the requirement $F_{\text{DR1}}\approx
F_{\text{DR2}}$ is a necessary but not a sufficient condition for the validity
of DR.
\begin{figure}[t]
\centerline{\resizebox{\hsize}{!}{\includegraphics{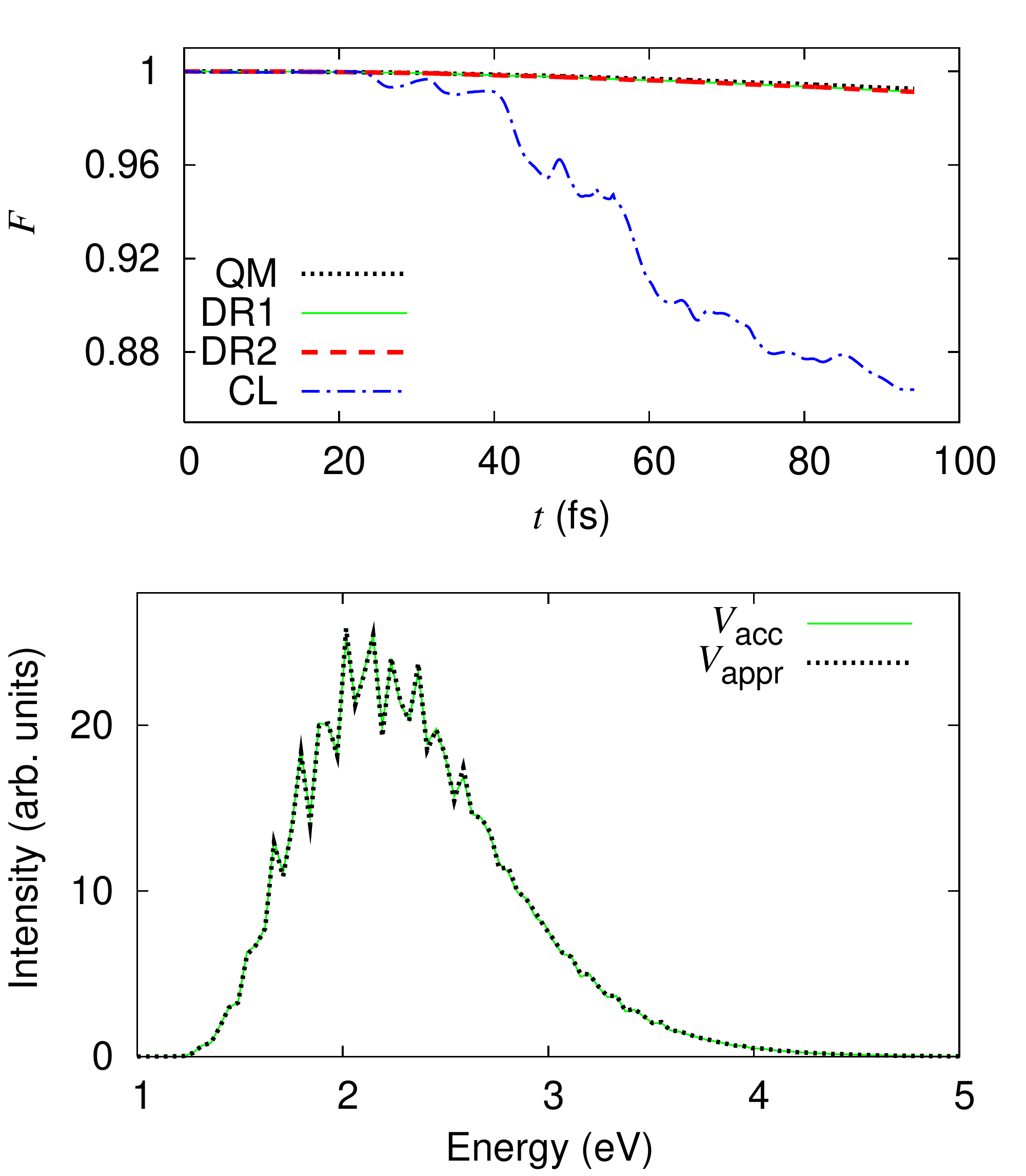}}}
\caption{Fidelity decay (top) and spectrum change (bottom) for
  $D_{e,\text{CO}}^{\prime} = D_{e,\text{CO}} + 0.02\operatorname{eV} =11.26\operatorname{eV}$.}
\end{figure}
\begin{figure}[t]
\centerline{\resizebox{\hsize}{!}{\includegraphics{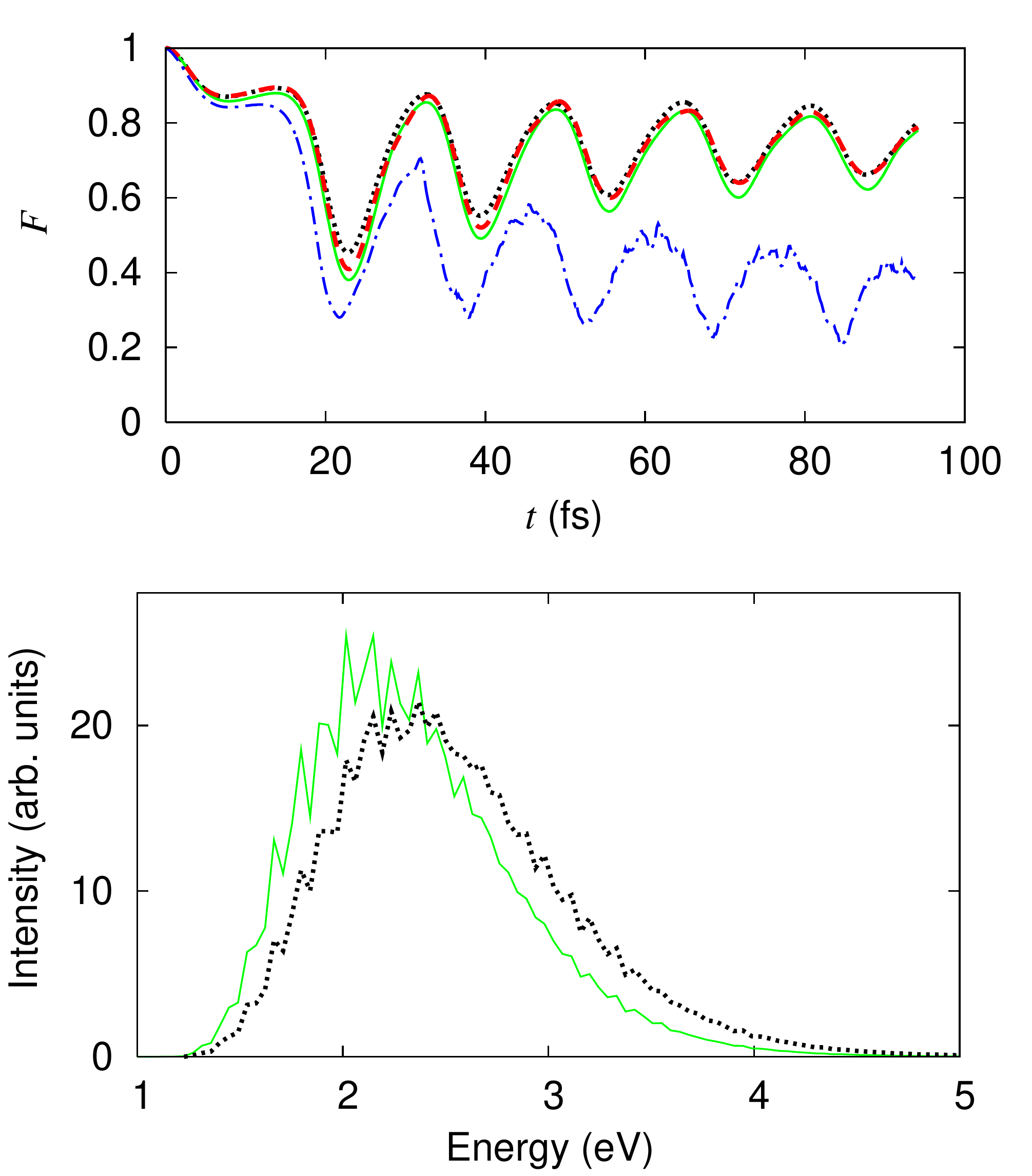}}}
\caption{Fidelity decay (top) and spectrum change (bottom) for
  $R_{e,\text{CO}}^{\prime} = R_{e,\text{CO}} + 0.02\,a_{0}
  =2.15\,a_{0}$. Notation as in Fig.~1.}
\end{figure}

One could object that since DR is an intrinsically SC
approximation, it might suffice to estimate fidelity classically. We explored
this idea by comparing quantum fidelity with its classical (CL) analog, called
classical fidelity \cite{prosen:2002a}, defined in our case by%
\begin{equation}
F_{\text{CL}}(t):=h^{d}\int dx\,\rho_{\text{CL,acc}}^{t}(x)\rho
_{\text{CL,appr}}^{t}(x),
\end{equation}
where $\rho_{\text{CL}}^{t}$ is the CL phase-space density evolved with
the indicated potential to time $t$. For initial Gaussian wave packets, $\rho_{\text{CL}}^{0}=\rho_{\text{W}}^{0}$. Unlike DR, classical fidelity
does not include any dynamical quantum effects, and below we show that
indeed $F_{\text{CL}}$ gives much worse results.

To show the feasibility of our method, we have applied it to the
photodissociation dynamics of a collinear carbon dioxide molecule, a model that had been studied extensively
by both quantum-dynamical \cite{kulander:1980,schinke:1990} and SC
methods \cite{vanvoorhis:2002}. Invoking the Franck-Condon principle,
photodissociation process is described by the quantum dynamics of the initial state
(vibrational ground state of the electronic ground state PES) on the
dissociative excited PES. 
One can obtain the photodissociation spectrum simply by taking the Fourier
transform of the autocorrelation function
$C(t):=\langle\psi(0)|\psi(t)\rangle$. In future applications, one will not be able to find $\psi
_{\text{acc}}(t)$ and $f_{\text{QM}}(t)$ due to the tremendous
computational expense. Here, in order to demonstrate the accuracy of the method, we
want to compare $f_{\text{DR}}(t)$ with $f_{\text{QM}}(t)$ and so we define
$V_{\text{acc}}:=V_{\text{LEPS}}$ to be the analytical LEPS potential for
CO$_{2}$ \cite{sato:1955,kulander:1980} and $V_{\text{appr}}:=V_{\text{LEPS}%
}^{\prime}$ to be the LEPS potential with one of the parameters perturbed.
Specifically, we change either the equilibrium bond length $R_{e,\text{CO}}$
or the bond dissociation energy $D_{e,\text{CO}}$ of the CO bond. We imagine
that we would like to obtain a spectrum corresponding to $V_{\text{acc}}$ but
can only afford quantum dynamics on $V_{\text{appr}}$. By estimating quantum
fidelity amplitude $f_{\text{QM}}$ by $f_{\text{DR}}$, we can determine
whether we can trust the spectrum computed using quantum dynamics on
$V_{\text{appr}}$. From another perspective, we can use $f_{\text{DR}}$ to evaluate how errors in experimental values of $R_{e,\text{CO}}$
and $D_{e,\text{CO}}$ affect the computed spectrum.

Figure 1 shows an example where $V_{\text{LEPS}}^{\prime}$ has a perturbed
bond dissociation energy, $D_{e,\text{CO}}^{\prime}=D_{e,\text{CO}}+0.02%
\operatorname{eV}%
=11.26%
\operatorname{eV}%
$. The figure shows that three approaches to compute fidelity (QM, DR1,
DR2) give very similar results. This turns out to be a small perturbation
since fidelity remains close to unity, $F_{\text{QM}}\geq0.99$. Therefore we
should be able to trust the spectrum computed using $\psi_{\text{appr}}(t)$.
This is justified in the bottom panel where spectra
corresponding to $V_{\text{acc}}$ and $V_{\text{appr}}$ prove
to be almost identical \endnote{Spectrum is significantly affected for $\Delta D_{e,\text{CO}} \gtrsim
  0.1 \operatorname{eV}.$}. Unlike DR, classical fidelity (CL) decays much
faster than its quantum analog. Judging by $F_{\text{CL}}$ only, one would
conclude incorrectly that the spectrum computed using $V_{\text{appr}}$ should
not be trusted. Partially constructive interference, captured by DR, prevents
quantum fidelity from decaying with the fast rate of classical fidelity decay.

Figure 2 shows an example where $V_{\text{LEPS}}^{\prime}$ has a perturbed
equilibrium bond length, $R_{e,\text{CO}}^{\prime}=R_{e,\text{CO}}%
+0.02\,a_{0}=2.15\,a_{0}$. The figure shows that $F_{\text{DR}}$ agrees
with $F_{\text{QM}}$ and even reproduces detailed oscillations of quantum
fidelity. This turns out to be a large perturbation since fidelity falls below
the value of $0.5$. We therefore expect the spectrum computed using
$\psi_{\text{appr}}(t)$ to differ significantly from the \textquotedblleft
true\textquotedblright\ spectrum computed using $\psi_{\text{acc}}(t)$.
Indeed, the bottom panel shows that the spectrum of $V_{\text{appr}}$ is
shifted and has different peak intensities than the spectrum of $V_{\text{acc}%
}$. Figure 2 also shows that for large $\Delta V$, classical fidelity starts
to behave similarly to quantum fidelity, but is still much worse than DR.

All calculations were performed for $600$ time steps
of $0.16%
\operatorname{fs}%
$ each. Converged quantum calculations required $n=1024$ points to discretize each
CO bond length from $0$ to $20\,a_{0}\approx10.6%
\operatorname{\mathring{A}}%
$, altogether using a $1024\times1024$ grid to represent the PES and the wave
function. The
DR calculation converged fully with $n_{\text{paths}} = 512$
trajectories (shown in Figs. 1 and 2), but a much smaller value, $n_{\text{paths}}
= 64$, already gives very accurate results sufficient for our
application (not shown). Both figures show fidelity $F$,
rather than fidelity amplitude $f$ which is a complex number containing more
information. If $F \ll 1$,
clearly our approximate dynamics is not sufficient. However, if $F\approx1$,
the spectrum could still be affected by a time-dependent phase of $f$. In such cases one should also examine fidelity amplitude.

Our fidelity calculation for photodissociation of CO$_{2}$ is to
our knowledge the first fidelity calculation for a realistic chemical system
and it is reassuring that DR remains valid \cite{gorin:2006}. Clearly, in
chemical physics one is interested in systems with many more than two
DOF. However, already in the simple CO$_{2}$ system, the DR calculation of
fidelity was more than $100$ times faster than the exact quantum
calculation. We expect that in larger systems, much larger speedups could be
achieved. 

The information required in a DR calculation is similar to information needed
in molecular dynamics (MD). Implementation of DR into any MD code
would require a single addition: calculation of the action $\Delta S$. But there is an important
difference between DR and MD: in DR, nuclear quantum effects are
included at least approximately, whereas in MD, even if ab initio electronic potential is
used, they are completely lost. This can be clearly seen in
Figs. 1 and 2 where $F_{\text{DR}}\approx F_{\text{QM}}\neq F_{\text{CL}}$
since $F_{\text{CL}}$ is basically a MD calculation of fidelity. With
little effort, MD codes that currently compute only classical nuclear
dynamics, could be used for evaluating accuracy of quantum dynamics on the
same PES. 

The method was designed with the goal of determining the accuracy of quantum
dynamics on an approximate PES compared to the exact dynamics. As mentioned
above, we do not know $V_{\text{exact}}$ and instead must use $V_{\text{acc}}%
$. We predict that if one could show $V_{\text{acc}}$ to be much closer to $V_{\text{exact}}$
than to $V_{\text{appr}}$, our estimate of fidelity would predict the
accuracy compared to the exact dynamics, and not just compared to the
dynamics on $V_{\text{acc}}$.

The DR approach is applicable to other types of perturbations of the Hamiltonian than
discrepancies in the PES. For example, DR could be used to evaluate
how laser pulse noise affects quantum control \cite{li:2002} or how
perturbations affect quantum computation. To conclude, we do not
claim to have found a fast way to do quantum dynamics on an accurate ab initio
potential. Instead we have found a promising method
to estimate the accuracy of quantum dynamics on an approximate potential.

This research was supported by the startup funding provided by \'{E}cole
Polytechnique F\'{e}d\'{e}rale de Lausanne. We thank Tom\'{a}\v{s} Zimmermann
for useful discussions.

\end{document}